\documentclass[]{rnaastex}

\usepackage{graphicx}
\usepackage[suffix=]{epstopdf}
\usepackage{natbib}
\usepackage{amsmath}
\usepackage{xspace}

\usepackage{hyperref}

\begin{document}



\newcommand{\kepler}{\emph{Kepler}}

\newcommand{\python}{{\tt PYTHON}}
\newcommand{\phasma}{{\tt phasma}}
\newcommand{\forecaster}{{\tt forecaster}}
\newcommand{\emcee}{{\tt emcee}}
\newcommand{\scipy}{{\tt scipy}}
\newcommand{\astropy}{{\tt astropy}}
\newcommand{\sinc}{\mathrm{sinc}}

\newcommand{\hardcore}{{\tt hardCORE}}

\title{TRAPPIST-1e HAS A LARGE IRON CORE}

\correspondingauthor{Gabrielle Suissa}
\email{ge2205@columbia.edu}

\author{Gabrielle Suissa \& David Kipping}
\affiliation{Department of Astronomy, Columbia University, 550 W 120th Street, New York NY}

\keywords{planets and satellites: detection}

\section{} 

The TRAPPIST-1 system provides an exquisite laboratory for understanding
exoplanetary atmospheres and interiors. Their mutual gravitational interactions
leads to transit timing variations, from which \citet{grimm:2018} recently
measured the planetary masses with precisions ranging from 5\% to 12\%. Combined
with $<5$\% radius measurements on each planet, TRAPPIST-1 provides a unique
opportunity to examine the range of permissible planetary interiors.

\citet{grimm:2018} used their new masses and radii and compared them to those
expected for planets comprised of pure silicate (no iron or volatiles). This
revealed that planets b, d, f, g and h likely contain volatile layers to
explain their properties, but c and e are compatible with being rocky. This is
an example of a boundary condition comparison, first described in
\citet{kipping:2013} in the context of planetary interiors, where the authors
show how a minimum envelope height can be derived by comparison to pure water
models. We briefly note that planets b through h all have a minimum envelope
height compatible with zero when applying the method of \citet{kipping:2013} to
the \citet{grimm:2018} masses and radii, to a confidence of $\geq 99.999$\%.

\citet{grimm:2018}'s inference that planets c and e are consistent with a
rock-iron composition is useful, but it is possible to go further and actually
quantify the minimum and maximum size of an iron core using boundary condition
arguments. Such an approach is laid out in our recent paper
\citet{suissa:2018}. In that work, we considered that the maximum core size
is found by solving when the mass and radius of the planet equals that of
an iron core surrounded by a light hydrogen/helium envelope. However, recent
atmospheric studies by \citet{dewit:2016,dewit:2018} exclude the possibility of
such envelopes for planets b through f. Accordingly, we updated our model,
\hardcore, such that the maximum core size corresponds to the next lightest
layer plausibly found around the core, a water layer (where as in our
original paper we use the \citet{zeng:2013} interior model).

Using the \citet{grimm:2018} posteriors, we are then able to derive a minimum
and maximum core size for each planet (using the original maximum formulation
for planets g and h). We find that the minimum core size is consistent with
zero for all of the planets except e. In particular, for planet c, unlike
the result of \citet{grimm:2018}, we find that the radius of the planet if
pure silicate would be $1.09_{-0.04}^{+0.04}$\,$R_{\oplus}$, which is
consistent with the observed radius of $1.09_{-0.03}^{+0.03}$\,$R_{\oplus}$.
The difference is likely a product of the different equations-of-state used in
each model, that of \citet{zeng:2013} used in our work, versus that of
\citet{connolly:2009} used by \citet{grimm:2018}. We find that the probability
of an iron core is modest at 57\% and thus ambiguity remains regarding c's
interior. For planet e, however, 99.3\% of the posterior samples are consistent
with a silicate-iron model indicating strong evidence for an iron core.
Substituting the silicate layer for water or volatile envelope naturally
increases this probability.

\begin{figure*}
\begin{center}
\includegraphics[width=15.5cm,angle=0,clip=true]{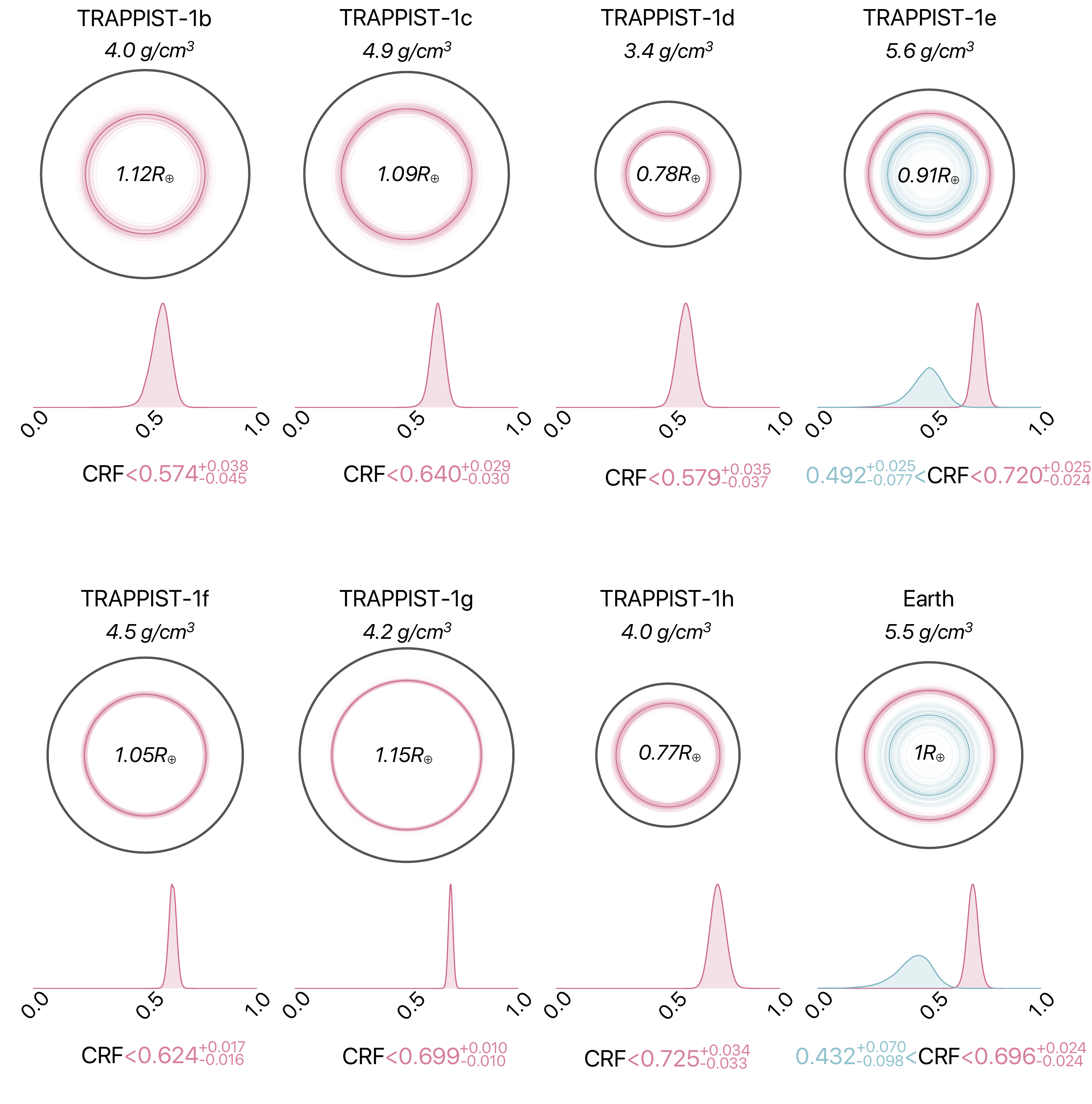}
\caption{\emph{
One hundred posterior samples of the maximum core radius fraction ($CRF$) of
each planet (with radii scaled correctly) are depicted with the thin red lines,
with the median shown using a thicker line. For planet e only, we find a
non-zero minimum CRF, which is shown in blue. The posteriors are depicted below
each planet as a histogram. Bottom-right panel shows the Earth if observed with
the same fractional uncertainty as TRAPPIST-1e as a point of reference.
}}
\label{fig:1}
\end{center}
\end{figure*}

Figure~\ref{fig:1} summarizes our findings, where we highlight that TRAPPIST-1e
has a minimum iron core size of $49.2_{-7.7}^{+6.2}$\%. This is remarkably
similar to that of Kepler-36b of $49.7_{-7.4}^{+6.7}$\% and both planets have
compositions entirely consistent with that of the Earth (for which we would
measure $\mathrm{CRF}>43$\%). This work demonstrates the power of boundary
conditions in making definitive statements regarding planetary interiors
without any assumptions of chemical relationship to the parent star.

\acknowledgments

DMK is supported by the Alfred P. Sloan Foundation.


\end{document}